%% file: arxiv.tex
\documentclass[onecolumn,a4paper]{IEEEtran}
\IEEEoverridecommandlockouts
\usepackage[left=1.49cm,right=1.49cm,top=1.9cm,bottom=4.1cm]{geometry}
\usepackage{mleftright}       
\mleftright                   

\usepackage{graphicx}         
\usepackage{booktabs} 

\usepackage[utf8]{inputenc}
\usepackage{amsmath,amssymb,amsfonts}
\usepackage{mathtools}
\usepackage{amsthm}
\usepackage{algorithmic}
\usepackage{ragged2e}
\usepackage{textcomp}
\usepackage{tikz}
\usepackage{caption}
\usepackage{cuted}
\usepackage{pgfgantt}
\usepackage{pdflscape}
\usepackage{pst-plot}
\usepackage{comment} 
\usepackage{cases}
\usepackage{nccfoots}
\usepackage{lineno}
\usepackage{graphicx}
\usepackage[caption=false,font=footnotesize]{subfig}
\usetikzlibrary{spy}
\usetikzlibrary{positioning,calc}
\usetikzlibrary{decorations.pathmorphing,calc,shapes,shapes.geometric,patterns}
\usetikzlibrary{shapes.multipart}
\usepackage{xfrac}
\usepackage{colortbl}
\usepackage{cancel} 
\usepackage{bbm}
\usetikzlibrary{arrows,positioning,calc,intersections}
\usetikzlibrary{datavisualization.formats.functions}
\def\BibTeX{{\rm B\kern-.05em{\sc i\kern-.025em b}\kern-.08em
    T\kern-.1667em\lower.7ex\hbox{E}\kern-.125emX}}
    
\usepackage{romannum}
\usepackage{pgfplots}
\usepgfplotslibrary{fillbetween}
\usetikzlibrary{arrows, decorations.markings}
\usetikzlibrary{arrows.meta}

\newtheorem{theorem}{Theorem}
\newtheorem*{theorem*}{Theorem}
\newtheorem{lemma}[theorem]{Lemma}

\newtheorem{definition}[theorem]{Definition}

\newtheorem{corollary}[theorem]{Corollary}
\newtheorem{remark}[theorem]{Remark}

\DeclarePairedDelimiterX{\infdivx}[2]{(}{)}{%
  #1\;\delimsize\|\;#2%
}

\pgfplotsset{compat=1.18}
\begin{document}
\title{Identification via Gaussian Multiple Access Channels in the Presence of Feedback} 



\author{
\IEEEauthorblockN{Yaning Zhao\IEEEauthorrefmark{1}\IEEEauthorrefmark{2}, Wafa Labidi \IEEEauthorrefmark{1}\IEEEauthorrefmark{2}\IEEEauthorrefmark{4}\thanks{\IEEEauthorrefmark{4}BMBF Research Hub 6G-life, Germany}, Holger Boche\IEEEauthorrefmark{1}\thanks{\IEEEauthorrefmark{3}Cyber Security in the Age of Large-Scale Adversaries–
Exzellenzcluster, Ruhr-Universit\"at Bochum, Germany}\IEEEauthorrefmark{4}\thanks{\IEEEauthorrefmark{6}{\color{black}{ Munich Center for Quantum Science and Technology (MCQST) }}}\IEEEauthorrefmark{7}\thanks{\IEEEauthorrefmark{7} {\color{black}{Munich Quantum Valley (MQV)}} }, Eduard Jorswieck\IEEEauthorrefmark{2}\IEEEauthorrefmark{5} \thanks{\IEEEauthorrefmark{5}BMBF Research Hub 6G-RIC, Germany}, and Christian Deppe\IEEEauthorrefmark{2}\IEEEauthorrefmark{4}\\} 
\IEEEauthorblockA{\IEEEauthorrefmark{1}Technical University of Munich,
\IEEEauthorrefmark{2}Technical University of Braunschweig\\
Email: yaning.zhao@tu-bs.de, wafa.labidi@tum.de, boche@tum.de, e.jorswieck@tu-bs.de, christian.deppe@tu-bs.de}
}

\maketitle


\begin{abstract}
We investigate message identification over a K-sender Gaussian multiple access channel (K-GMAC). Unlike conventional Shannon transmission codes, the size of randomized identification (ID) codes experiences a doubly exponential growth in the code length. Improvements in the ID approach can be attained through additional resources such as quantum entanglement, common randomness (CR), and feedback. It has been demonstrated that an infinite capacity can be attained for a single-user Gaussian channel with noiseless feedback, irrespective of the chosen rate scaling. We establish the capacity region of both the K-sender Gaussian multiple access channel (K-GMAC) and the K-sender state-dependent Gaussian multiple access channel (K-SD-GMAC) when strictly causal noiseless feedback is available.
\end{abstract}

\section{Introduction}
\label{sec:introduction}
The first systematic study of the transmission problem was reported by Shannon \cite{shannon1948mathematical} in 1948. His finding reveals that the maximal size of reliably transmitted messages grows exponentially with the block length. 
In recent years, various applications in 6G communication systems \cite{fettweis20226g,cabrera20216g}, such as machine-to-machine systems \cite{boche2018secure}, digital watermarking \cite{moulin2001role,ahlswede2006watermarking,steinberg2001identification}, and Industry 4.0 \cite{lu2017industry} emphasize the need for higher capacity, lower latency, and improved data security \cite{Schwenteck2023}. Ahlswede and Dueck \cite{ahlswede1989identification}, introduced the ID scheme building on Jaja's work \cite{ja1985identification}, which proves to be more efficient than transmission in many scenarios. In the ID problem, the receiver only needs to decide if the sender has transmitted the message in which the receiver is interested. Of course, the sender does not know which message is actually interesting to the receiver. Using randomized ID coding schemes, the size of ID messages (also called identities) can grow doubly exponentially with the block length \cite{ahlswede1989identification}.
Recent research has shown the seamless integration of information-theoretic security into identification protocols \cite{ahlswede1995new,labidi2020secure}. Advancements in ID can tap into a diverse array of resources, including quantum entanglement \cite{boche2019secure,pereg2022identification}, common randomness (CR) \cite{ezzine2024common,ezzine2021common,labidi2022common}, and feedback \cite{ahlswede1989identificationfeedback}. Quantum entanglement often demonstrates superior effectiveness compared to CR.

Feedback sets apart transmission and identification schemes. Feedback does not enhance transmission capacity for discrete memoryless channels (DMCs) \cite{wolfowitz2012coding,cover1989gaussian}. However, feedback can increase the ID capacity \cite{ahlswede1989identificationfeedback}. It enables the use of randomized ID codes and allows the ID capacity of DMCs to grow doubly exponentially with code length \cite{ahlswede1989identificationfeedback}. 
Many studies have explored the ID over continuous channels \cite{koga2002information,burnashev1999method,salariseddigh2021deterministicfading}. It has been shown that infinite CR can be generated from Gaussian source \cite{ezzine2020common}. The identification with feedback (IDF) problem via single-user Gaussian channels has been explored in \cite{labidi2020secure,labidi2021identification,9716126}, demonstrating the achievability of infinite capacity regardless of the chosen rate scaling.

The MAC is a communication model designed to facilitating simultaneous data transmission from multiple senders to the receiver. The transmission problem via MAC was initially characterized by Liao \cite{liao1972multiple}, and further insights were contributed in \cite{el2011network}. Noiseless feedback can increase the transmission capacity of MAC \cite{gaarder1975capacity}. Previous studies have established the upper bound of transmission capacity for Gaussian MAC with feedback \cite{cover1999elements,thomas1987feedback}, showing that the total capacity of any MAC with additive Gaussian white noise (AWGN) can at most be doubled by feedback.
The deterministic ID capacity of discrete memoryless MAC (DM-MAC) is explored in \cite{rosenberger2023deterministic}, while \cite{ahlswede2008general} investigates randomized ID capacity. Quantum entanglement's enhancement of randomized ID via DM-MAC is shown in \cite{diadamo2019simultaneous}, with further advantages of perfect feedback discussed in \cite{ahlswede1971multi}. To the best of our knowledge, no study has yet investigated ID via GMAC in the presence of feedback. We establish the IDF capacity region of K-GMAC and K-SD-GMAC, where each sender can achieve an infinite double-exponential growth rate. We demonstrate achievability by providing a deterministic coding scheme such each sender can achieve an infinite rate. Furthermore, similar to single-user Gaussian channels, this result holds irrespective of the definition of rates.

In this paper, as well as in the referenced literature (with the exception of \cite{ahlswede1995new, Kleinewachter1999}), it is assumed that the feedback is noise-free. To the best of our knowledge, the scenario of noisy feedback has been explored exclusively in \cite{ahlswede1995new, Kleinewachter1999}. In that study, even when considering identification over a point-to-point channel with noisy feedback, only upper and lower bounds for the identification capacity were established.

\textit{Outline}: The remainder of the paper is structured as follows. In Section \ref{sec:preliminaries}, we review previous results on the IDF problem via discrete and Gaussian single-user channels. In Section \ref{sec:models&results}, we introduce our system model and present the main results. Section \ref{sec:proof} is dedicated to establishing a proof of the infinite ID capacity region for K-GMAC and K-SD-GMAC. Finally, Section \ref{sec:conclusion} concludes the paper.

\section{Preliminaries}
\label{sec:preliminaries}
In this section, we recall the previous results in \cite{labidi2021identification} on IDF via single-user Gaussian channel.
\begin{figure}\begin{center}
    \hspace{-2em}
    \input{Figures/IDFsingle}
    \caption{IDF via single-user Gaussian channel with feedback}
    \vspace{-0.5em}
    \label{fig:GaussianChannel}
\end{center}\end{figure}
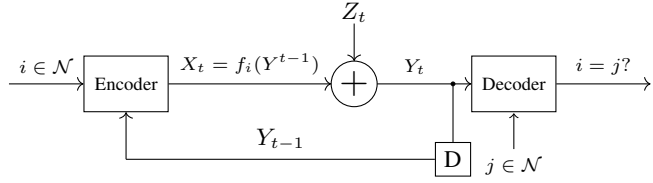
Consider the IDF problem via a single-user discrete-time Gaussian channel $W_{\sigma^2}$ as depicted in Fig. \ref{fig:GaussianChannel}. For all $t=1,\cdots,n$, the channel output $Y_t$ is given by
\begin{align*}
    Y_t=X_t+Z_t,
\end{align*}
where $X_t$ is the channel input and $Z_t$ is AWGN following the normal distribution with mean $0$ and variance $\sigma^2$, i.e., $Z_t\sim\mathcal{N}\left(0,\sigma^2\right)$.
The feedback encoding function for a Gaussian channel with average power constraints is defined as follows.
\begin{definition}\label{def:f} Let $\mathcal N$ be a message set and $i\in\mathcal N$. Feedback encoding functions $\boldsymbol{f}_i$ for Gaussian channels are vector-valued functions defined as follows:
    \begin{align*}
    \boldsymbol{f}_{i}=\left[f^1_{i},\cdots,f^n_{i}\right],
\end{align*}
where $f^1_{i}\in\mathbb{R}$, and for $t=\left\{2,\cdots,n\right\}$, $f^t_{i}:\mathbb{R}^{t-1}\mapsto \mathbb{R}$. 

These functions satisfy for $P_{total}\in\mathbb{R}^+$ the following average power constraints:
\begin{align*}
    \sum_{t=1}^n \left(f^t_{i}\right)^2\le n\cdot P_{total}, \quad \forall i\in\mathcal{N}.
\end{align*}
We denote the set of all such functions with length $n$ as $\bar{\mathcal{F}}_n$.
Similarly, we denote the set of all the functions that satisfy the following peak power constraints $P_{peak}\in\mathbb{R}^+$
\begin{align*}
    |f_{i,t}|\le P_{peak},\quad t=1,\cdots,n
\end{align*}
as $\hat{\mathcal{F}}_n$.
\end{definition}
\begin{definition} An deterministic $(n,N,\lambda)$ IDF code with $\lambda\in\left(0,\frac{1}{2}\right)$ for a Gaussian channel $W_{\sigma^2}$ is a system $\left\{\left(\boldsymbol{f}_i,\mathcal{D}_i\right)|i\in\mathcal{N}\right\}$ with
\begin{align*}
   \boldsymbol{f}_i\in \bar{\mathcal{F}}_n, \quad \mathcal{D}_i\subset \mathbb{R}^n,\quad \forall i\in\mathcal{N},
\end{align*}
such that
\begin{align*}
    P_{e,1}(i)&\triangleq W_{\sigma^2}^n(\mathcal{D}_i^c|\boldsymbol{f_i})\le \lambda,\quad \forall i\in\mathcal{N},\\
    P_{e,2}(i,j)&\triangleq W_{\sigma^2}^n(\mathcal{D}_{\tilde{i}}|\boldsymbol{f_i})\le \lambda,\quad \forall i,\tilde{i}\in\mathcal{N},i\ne\tilde{i}.
\end{align*}
\end{definition}
The deterministic IDF capacity of a single-user Gaussian channel is as follows.
\begin{theorem}\cite{labidi2021identification} 
    Let $P>0$. Then there exists for all $R>0$ a blocklength $n_0$ such that for every $n\ge n_0$ there exists a deterministic IDF code for $W_{\sigma^2}$ of blocklength $n$ with $N=2^{2^{nR}}$ identities and with $\lambda\in\left(0,\frac{1}{2}\right)$, i.e., $C(W_{\sigma^2},P)=+\infty$.
\end{theorem}
\section{System Model and Main Results}
\label{sec:models&results}
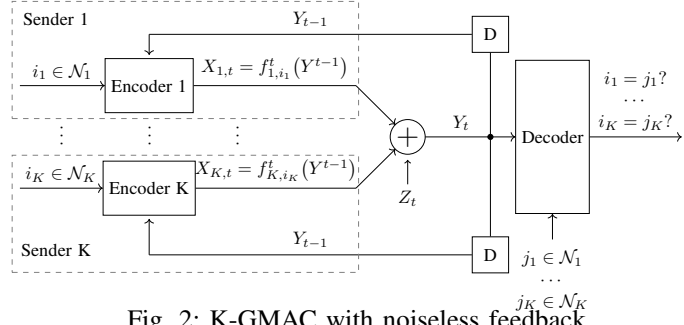
\begin{figure}\begin{center}
    \hspace{-2.5em}
    \input{Figures/GAMC}
    \captionsetup{margin=-5pt}
    \vspace{-1em}
    \caption{K-GMAC with noiseless feedback}
    \vspace{-1em}
    \label{fig:GMAC}
\end{center}\end{figure}

Consider a K-GMAC with strictly causal noiseless feedback as depicted in Fig. \ref{fig:GMAC}. Each sender $k\in\{1,\cdots,K\}\triangleq\mathcal{K}$ aims to send an n-length sequence $X_k^n=\left(X_{k,1}\cdots,X_{k,n}\right)\in\mathbb{R}^n$ over the forward AWGN channel denoted by $W_{\sigma^2}$. The channel output $Y_t$ can be expressed as:
\begin{align*}
    Y_t=\sum_{k=1}^{K}{X_{k,t}}+Z_t,\quad t=1,\cdots,n,
\end{align*}
where $Z_1,\cdots,Z_n$ are independent and identically distributed (i.i.d.) following a normal distribution with zero mean and variance $\sigma^2$, i.e., $Z_t\sim \mathcal{N}(0,\sigma^2)$, for all $t=1,\cdots,n$. Given $j_k\in\mathcal{N}_k$, the decoder wants to determine whether $i_k=j_k$ holds separately for each $k\in\mathcal{K}$.

Each encoder $k\in\mathcal{K}$ independently encodes an identity $i_k\in\mathcal{N}_k$ w.r.t. a strictly causal noiseless feedback sequence $y^{t-1}$ to a codeword $x_k^n$. The feedback encoding functions $\boldsymbol{f}_{k,i_k}\in\bar{\mathcal{F}}_n$ are defined in the same manner as in Definition~\ref{def:f}.
We denote the identity tuple as $\boldsymbol{i}=\left(i_1,\cdots,i_K\right)$, the identity set tuple as $\boldsymbol{\mathcal{N}}=\left(\mathcal{N}_1,\cdots,\mathcal{N}_K\right)$, the cardinality tuple as $\boldsymbol{N}=(N_1,\cdots,N_2)$, the feedback encoding function tuple as $\mathbf{f}_{\boldsymbol{i}}=\left(\boldsymbol{f}_{1,i_1},\cdots,\boldsymbol{f}_{K,i_K}\right)$, and the codeword tuple as $\boldsymbol{X}^n=\left(X_1^n,\cdots,X_K^n\right)=\boldsymbol{f}_{\boldsymbol{i}}\left( Y^n \right)$. In the following, we define the deterministic IDF code for a K-GMAC.
\begin{definition}
    An $\left(n,\boldsymbol{N},\lambda\right)$ deterministic IDF code with $\lambda\in\left(0,\frac{1}{2}\right)$ for $W_{\sigma^2}$ is a system $\left\{\left(\boldsymbol{f}_{\boldsymbol{i}},\mathcal{D}_{\boldsymbol{i}}\right)|\boldsymbol{i}\in\boldsymbol{\mathcal{N}}\right\}$, where
    \begin{align*}
        \boldsymbol{f}_{\boldsymbol{i}}\in\bar{\mathcal{F}}_n^{K},\quad \mathcal{D}_{\boldsymbol{i}}\subset \mathcal{Y}^n.
    \end{align*}
    The probabilities of type I error and type II error satisfy
    \begin{align*}
        P_{e,1}\left(\boldsymbol{i}\right)&\triangleq W_{\sigma^2}\left(\mathcal{D}^c_{\boldsymbol{i}}|\boldsymbol{f}_{\boldsymbol{i}}\right)\le \lambda,\quad \forall \boldsymbol{i}\in\boldsymbol{\mathcal{N}},\\
        P_{e,2}\left(\boldsymbol{i},\tilde{\boldsymbol{i}}\right)&\triangleq W_{\sigma^2}(\mathcal{D}_{\boldsymbol{\tilde{i}}}|\boldsymbol{f}_{\boldsymbol{i}})\le \lambda,\quad \forall \boldsymbol{i},\tilde{\boldsymbol{i}}\in\boldsymbol{\mathcal{N}},\tilde{\boldsymbol{i}}\ne \boldsymbol{i}.
    \end{align*}
\end{definition}
We first consider the rates defined under double-exponential scaling, i.e., $R_k=\frac{\log{\log{N_k}}}{n}$, for all $k\in\mathcal{K}$. Denote the rate tuple as $\boldsymbol{R}=2^{2^{\boldsymbol{N}}}$.
\begin{definition}

    \begin{enumerate}
        \item A rate tuple $\boldsymbol{R}=\left(R_1,\cdots,R_K\right)$ is said to be achievable, if for $\lambda\in\left(0,\frac{1}{2}\right)$, there exists an $n_0(\lambda)$, such that for all $n\ge n_0(\lambda)$, there exists an $\left(n,\boldsymbol{N}=2^{2^{n\boldsymbol{R}}},\lambda\right)$ IDF code.
        \vspace{1em}
        \item The IDF capacity region is defined as the closure of all achievable $\boldsymbol{R}$.
    \end{enumerate}
\end{definition}
\begin{theorem}
\label{thm:GMAC}
If $P_{total}>0$, then the IDF capacity region $\mathcal{C}$ of K-GMAC $W_{\sigma^2}$ is given by
\begin{align*}
    \mathcal{C}\left(W_{\sigma^2},P_{total}\right)=\left\{\boldsymbol{R}:R_k< +\infty, \quad \forall k\in\mathcal{K}\right\}.
\end{align*}
\end{theorem}
\begin{figure}\begin{center}
    \hspace{-2em}
    \input{Figures/SDGMAC}
    \vspace{-1em}
    \caption{K-SD-GMAC with noiseless feedback}
    \label{fig:SD-GMAC}
    \vspace{-1em}
\end{center}\end{figure}
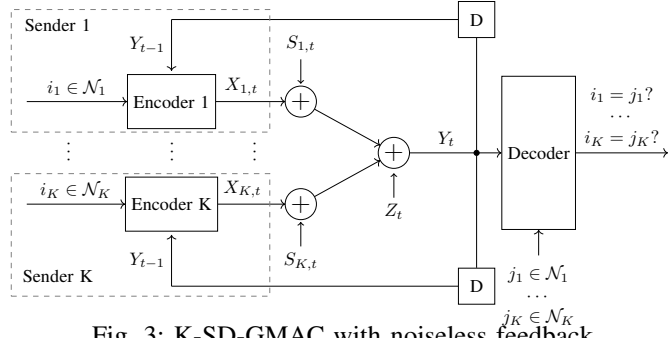
Next, we extend our model to K-SD-GMAC as illustrated in Fig.\ref{fig:SD-GMAC}. For each sender $k\in\mathcal{K}$, we introduce a channel state $S_{k}$. We assume the state vector $\boldsymbol{S}_t=(S_{1,t},\cdots,S_{K,t})^T$ is i.i.d. following a multivariate normal distribution with mean vector $\boldsymbol{\mu}$ and covariance matrix $\boldsymbol{\Sigma}$, i.e., $\boldsymbol{S}_t\sim \mathcal{N}_K\left(\boldsymbol{\mu},\boldsymbol{\Sigma}\right)$, where the probability density function (PDF) of $\boldsymbol{S}$ is denoted as $p_{\boldsymbol{S}}(\boldsymbol{s})$. We represent the channel transition probability of K-SD-GMAC as $W_{\sigma^2,S}$. The channel output $Y_t$ is given by
\begin{align*}
    Y_t=\sum_{k=1}^KX_{k,t}+\sum_{k=1}^K S_{k,t}+Z_t, \quad t=1,\cdots,n.
\end{align*}

\begin{definition}
    An $\left(n,\boldsymbol{N},\lambda\right)$ deterministic IDF code with $\lambda\in\left(0,\frac{1}{2}\right)$ for $W_{\sigma^2,S}$ is a system $\left\{\left(\boldsymbol{f}_{\boldsymbol{i}},\mathcal{D}_{\boldsymbol{i}}\right)|\boldsymbol{i}\in\boldsymbol{\mathcal{N}}\right\}$, where
    \begin{align*}
        \boldsymbol{f}_{\boldsymbol{i}}\in\bar{\mathcal{F}}_n^{K},\quad \mathcal{D}_{\boldsymbol{i}}\subset \mathcal{Y}^n.
    \end{align*}
    The probabilities of type I error and type II error satisfy
    \begin{align*}
        P_{e,1}\left(\boldsymbol{i}\right)&\triangleq{\int_{\boldsymbol{s}}} p_{\boldsymbol{S}}(\boldsymbol{s})W_{\sigma^2,S}(\mathcal{D}^c_{\boldsymbol{i}}|\boldsymbol{f}_{\boldsymbol{i}},\boldsymbol{s})d\boldsymbol{s}\le \lambda,\\ &\quad\quad\quad\quad\quad\quad\quad\quad\quad\quad\quad\quad\quad\quad\forall \boldsymbol{i}\in\boldsymbol{\mathcal{N}},\\
        P_{e,2}\left(\boldsymbol{i},\tilde{\boldsymbol{i}}\right)&\triangleq{\int_{\boldsymbol{s}}}p_{\boldsymbol{S}}(\boldsymbol{s})W_{\sigma^2,S}(\mathcal{D}_{\boldsymbol{i}}|\boldsymbol{f}_{\boldsymbol{i}},\boldsymbol{s})d\boldsymbol{s}\le \lambda,\\
        &\quad\quad\quad\quad\quad\quad\quad\quad\quad\quad\quad\quad\quad\quad\forall \boldsymbol{i},\tilde{\boldsymbol{i}}\in\boldsymbol{\mathcal{N}},\tilde{\boldsymbol{i}}\ne \boldsymbol{i}.
    \end{align*}
\end{definition}
The IDF capacity region under double-exponential scaling of K-SD-GMAC is as follows.
\begin{theorem}
\label{thm:SD-GMAC}
If $P_{total}>0$, then the IDF capacity region of K-SD-GMAC $W_{\sigma^2,S}$ is given by
\begin{align*}
    \mathcal{C}(W_{\sigma^2,S},P_{total})=\left\{\boldsymbol{R}:R_k< +\infty, \quad \forall k\in\mathcal{K}\right\}.
\end{align*}
\end{theorem}

\begin{remark}
    Theorem \ref{thm:GMAC} and Theorem \ref{thm:SD-GMAC} can be extended to the IDF  under peak power constraints $P_{peak}$ by choosing the feedback encoding function $\boldsymbol{f}_{\boldsymbol{i}}\in\hat{\mathcal{F}}_n^{K}$.
\end{remark}

\section{Proof}
\label{sec:proof}
In this section, we provide the proof of Theorem \ref{thm:GMAC} and Theorem \ref{thm:SD-GMAC} by presenting a coding scheme for K-GMAC and K-SD-GMAC, ensuring that each sender can achieve infinite capacity under average power constraints. This construction enables the generation of infinite common randomness shared among the encoders and the decoder, thereby allowing each sender to achieve an infinite rate regardless of the chosen scaling function $\varphi$. Given that the capacity is infinite, the necessity of a converse proof is obviated.
\subsection{Common Randomness Generation}
The generation of common randomness follows the steps outlined in \cite{labidi2021identification}. Consider any random variable (RV) $Y$ that follows a normal distribution with mean $\mu_Y$ and variance $\sigma_Y^2$. We can standardize $Y$ by defining $\bar{Y}=\frac{Y-\mu_Y}{\sigma_Y}$, where $\bar{Y}\sim \mathcal{N}\left(0,1\right)$. The cumulative distribution function (CDF) $\Phi(\cdot)$ of $\bar{Y}$ can be expressed as $F_{\bar{Y}}(\bar{y})=Pr\left[\bar{Y}\le \bar {y}\right]=\Phi(\bar{y})$. Introduce a new RV $\tilde{y}=\Phi(\bar{y})$, where
\begin{align*}
    F_{\tilde{Y}}(\tilde{y})
    &=Pr\left[\tilde{Y}\le\tilde{y}\right]\nonumber\\
    &=Pr\left[\Phi(\bar{Y})\le \tilde{y}\right]\nonumber\\
    &=Pr\left[\bar{Y}\le\Phi^{-1}(\tilde{y})\right]\nonumber\\
    &=\Phi\left(\Phi^{-1}\left(\tilde{y}\right)\right)\nonumber\\
    &=\tilde{y}.
\end{align*}
Consequently, we establish that $\tilde{Y}$ follows a uniform distribution $\mathcal{U}\left[0,1\right]$. To discretize $\tilde{Y}$, we proceed as follows. Let $\mathcal{L}=\left\{1,\cdots,L\right\}$ be set of positive integers, and define
\begin{align*}
\tilde{y}_l=\frac{l}{L},\quad 0\le l\le L.
\end{align*}
We partition the interval $\left[0,1\right]$ into $L$ segments using the endpoints $\tilde{y}_l$, i.e.,
\begin{align*}
\left[0,1\right]=\left[\tilde{y}_0,\tilde{y}_1\right]\cup\cdots\cup\left(\tilde{y}_{L-1},\tilde{y}_{L}\right].
\end{align*}
The channel output $Y$ is thereby partitioned into $L$ segments using endpoints $\sigma\cdot\phi^{-1}(\tilde{y}_l)$, i.e.,
\begin{align*}
\mathbb{R}
&=\left(\sigma_Y\phi^{-1}(\tilde{y_0})=-\infty,\sigma_Y\phi^{-1}(\tilde{y}_1)\right]\cup\cdots \\
&\quad \cdots\cup\left(\sigma_Y\phi^{-1}(\tilde{y}_{L-1}),\sigma_Y\phi^{-1}(\tilde{y}_L)=+\infty\right).
\end{align*}

Next, we define a function $u:\mathbb{R}\mapsto \mathcal{L}$, which maps $Y$ to a discrete uniformly distributed RV $U$ on the set $\mathcal{L}=\left\{1,\cdots,L\right\}$, defined by
\begin{align*}
U&=u(Y)=l,\quad \text{if } \sigma\phi^{-1}\left(\tilde{y}_{l-1}\right)<Y\le \sigma\phi^{-1}\left(\tilde{y}_{l}\right), \quad \forall l\in\mathcal{L}.
\end{align*}
The probability mass function (PMF) of $U$ is given by:
\begin{align*}
Pr\left[U=l\right]=\frac{1}{L}, \quad \forall l\in\mathcal{L}.
\end{align*}
Hence, any normally distributed RV $Y$ can be converted to a discrete uniformly distributed RV $U$ with zero error and without incurring additional costs during the uniformity and discretization process. Furthermore, as the interval $[0,1]$ is continuous, we can acquire as many $L$ samples as necessary, thereby generating common randomness of arbitrary size.
\subsection{Proof of Theorem \ref{thm:GMAC}}
In this section, we present the proof of Theorem \ref{thm:GMAC}. Consider an $\left(m,\boldsymbol{N}=2^{2^{m\boldsymbol{R}}}=\left(2^{2^{nR_1}},\cdots,2^{2^{nR_K}}\right),\lambda\right)$ deterministic IDF code $\left\{\left(\boldsymbol{f}_{\boldsymbol{i}},\mathcal{D}_{\boldsymbol{i}}\right)|\boldsymbol{i}\in\boldsymbol{\mathcal{N}}\right\}$ for K-GMAC, where the code length $m=n+1$. 

Initially, symbols $x_{k,1}=0$ are sent from each sender $k\in\mathcal{K}$ to generate the common randomness. We obtain $Y=Z\sim\mathcal{N}(0,\sigma^2)$, which is shared by both encoders (through the noiseless feedback links) and the decoder (through the forward channel). We convert $Y$ to $U=u(Y)\sim \mathcal{U}\left\{0,L\right\}$ as previously discussed.

Subsequently, we construct a tuple of $K$ families of functions, denoted as $\left\{\boldsymbol{F}_{\boldsymbol{i}}|\boldsymbol{i}\in\boldsymbol{\mathcal{N}}\right\}$, where $\boldsymbol{F}_{\boldsymbol{i}}=\left(F_{1,i_1},\cdots,F_{K,i_K}\right)$. Each function $F_{k,i_k}:\mathcal{L}\mapsto \mathcal{M}_k=\left\{1,\cdots,M_k\right\}$ corresponds to an identity $i_k\in\mathcal{N}_k$, mapping each element $u=u(y)\in\mathcal{L}$ to an integer $w_k=F_{k,i_k}(u)\in\mathcal{M}_k$. We uniformly randomly select the output of these mappings, i.e.,
\begin{align*}
    Pr\left[F_{k,i_k}(u)=w_k\right]=\frac{1}{M_k},\quad \forall w_k\in\mathcal{M}_k, \quad \forall k\in\mathcal{K}.
\end{align*}
We use $\boldsymbol{F}_{\boldsymbol{i}}(u)=\left(F_{1,i_1}(u),\cdots,F_{K,i_K}(u)\right)$ to denote the results of these functions. They are known to both the encoders and the decoder. We employ an $\left(n,\boldsymbol{M},2^{-n\delta}\right)$ standard  transmission code $\mathcal{C}'=\left\{\left(\boldsymbol{c}({\boldsymbol{F}_{\boldsymbol{i}}}(u)),\mathcal{D}'_{\boldsymbol{F}_{\boldsymbol{i}}(u)}\right)\right\}$ with average power constraint parameter $P$ to transmit $\boldsymbol{F}_{\boldsymbol{i}}(u)$. Here, $\boldsymbol{M}=\left(M_1,\cdots,M_K\right)$ denotes the tuple of the code sizes and $\boldsymbol{c}(\boldsymbol{F}_{\boldsymbol{i}})=\left(\boldsymbol{c}_1(F_{1,i_1}),\cdots,\boldsymbol{c}_K(F_{K,i_K})\right)$ denotes the tuple of the codewords.
For simplification, we use $\boldsymbol{c}$ to denote $\boldsymbol{c}(\boldsymbol{F_i}(l))$.

In conclusion, our deterministic code $\left\{\left(\boldsymbol{f}_{\boldsymbol{i}},\mathcal{D}_{\boldsymbol{i}}\right)|\boldsymbol{i}\in\boldsymbol{\mathcal{N}}\right\}$ is defined as follows:
\begin{align*}
    \boldsymbol{f}_{\boldsymbol{i}}&=\left(\boldsymbol{f}_{1,k_1},\cdots,\boldsymbol{f}_{K,i_K}\right), \quad \boldsymbol{f}_{k,i_k}=\left[0,\boldsymbol{c}_k(F_{k,i_k})\right],\\
    \mathcal{D}_{\boldsymbol{i}}&=\bigcup_{y\in\mathbb{R}}y\times\mathcal{D}'_{\boldsymbol{F}_{\boldsymbol{i}}(u(y))},\quad \forall \boldsymbol{i}\in\boldsymbol{\mathcal{N}}.
\end{align*}

For all $\boldsymbol{i}\in\boldsymbol{\mathcal{N}}$, the type I error can be bounded by
\begin{align*}
    P_{e,1}(\boldsymbol{i})
    &=W^m_{\sigma^2}\left(\left(\bigcup_{y\in\mathbb{R}}y\times\mathcal{D}'_{\boldsymbol{F}_{\boldsymbol{i}}(u(y))}\right)^c\Bigg|\boldsymbol{f}_{\boldsymbol{i}}\right)\\
    &\le \sum_{l\in\mathcal{L}}Pr[U=l] \mathbb{E}\left[W^n_{\sigma^2}\left(\left(\mathcal{D}'_{\boldsymbol{F}_{\boldsymbol{i}}(l)}\right)^c\big|\boldsymbol{c}\right)\big|U=l\right] \\
    &\overset{(a)}{\le} 2^{-n\delta}\\
    &=\circ(n),
\end{align*}
where (a) follows the definition of transmission code $C'$.
Next, we analyze the type II error probability $P_{e,2}(\boldsymbol{i},\boldsymbol{\tilde{i}})$ with $\boldsymbol{i}\ne \boldsymbol{\tilde{i}}$. We have
\begin{align*}
    P_{e,2}(\boldsymbol{i},\boldsymbol{\tilde{i}})
    &\le \max_{k\in\mathcal{K}}P^k_{e,2}(i_k,\tilde{i}_k),
\end{align*}
where $P^k_{e,2}(i_k,\tilde{i}_k)=W_{\sigma^2}(\mathcal{D}_{\boldsymbol{\tilde{i}_k}}|\boldsymbol{f}_{\boldsymbol{i}})$ 
with $\boldsymbol{\tilde{i}_k}=\left(i'_1,\cdots,\tilde{i}_k,\cdots,i'_K\right)$, for all $\boldsymbol{\tilde{i}_k}\in\boldsymbol{\mathcal{N}}$ and $i_k\ne\tilde{i}_k$.

Without the loss of generality, we examine the type II error probability $P^1_{e,2}$ for the identity set $\mathcal{N}_1$.
\begin{align*}
    P^1_{e,2}(i_1,\tilde{i}_1)
    &=W^m_{\sigma^2}\left(\bigcup_{y\in\mathbb{R}}y\times\mathcal{D}'_{\boldsymbol{F}_{\boldsymbol{\tilde{i}_1}}(l)}\Big|\boldsymbol{f}_{\boldsymbol{i}}\right)\\
    &=\sum_{l=1}^L Pr[U=l]\mathbb{E}\left[W^n_{\sigma^2}\left(\mathcal{D}'_{\boldsymbol{F}_{\boldsymbol{\tilde{i}_1}}(l)}\big|\boldsymbol{c}\right)\big|U=l\right]
\end{align*}

We partition $\mathcal{L}$ into four different sets as follows.
\begin{align*}
    F_{i_1}\cap F_{i'_1}&:= \left\{l\in\mathcal{L}|F_{1,i_1}(l)=F_{1,i'_1}(l)\right\},\\
    F_{i/i_1}\cap F_{i'/i'_1}&:= \left\{l\in\mathcal{L}|\boldsymbol{F}_{\boldsymbol{i/i_1}}(l)=\boldsymbol{F}_{\boldsymbol{i'/i'_1}}(l)\right\},\\
    F_{i_1}- F_{i'_1}&:= \left\{l\in\mathcal{L}|F_{1,i_1}(l)\ne F_{1,i'_1}(l)\right\},\\
    F_{i/i_1}- F_{i'/i'_1}&:= \left\{l\in\mathcal{L}|\boldsymbol{F}_{\boldsymbol{i/i_1}}(l)\ne\boldsymbol{F}_{\boldsymbol{i'/i'_1}}(l)\right\},
\end{align*}
where $\boldsymbol{i/i_1}=(i_2,\cdots,i_K)$ and $\boldsymbol{i'/i'_1}=(i'_2,\cdots,i'_K)$.

Then the type II error can be upper-bounded by
\begin{align*}
    &P^1_{e,2}(\boldsymbol{i},\boldsymbol{\tilde{i}_1})\\
    &=\sum_{l=1}^L Pr[U=l]\mathbb{E}\left[W^n_{\sigma^2}\left(\mathcal{D}'_{\boldsymbol{F}_{\boldsymbol{\tilde{i}_1}}(l)}\big|\boldsymbol{c}\right)\big|U=l\right]\\
    &= \sum_{l\in \{\{F_{i_1}\cap F_{\tilde{i}_1}\}\cap \{F_{i/i_1}\cap F_{i'/i'_1}\}\}}\frac{1}{L}\mathbb{E}\left[W^n_{\sigma^2}\left(\mathcal{D}'_{\boldsymbol{F}_{\boldsymbol{i}}(l)}\big|\boldsymbol{c}\right)\big|U=l\right]\\
    &+ \sum_{l\in \{\{F_{i_1}- F_{\tilde{i}_1}\}\cup \{F_{i/i_1}- F_{i'/i'_1}\}\}}\frac{1}{L}\mathbb{E}\left[W^n_{\sigma^2}\left(\mathcal{D}'_{\boldsymbol{F}_{\boldsymbol{\tilde{i}_1}}(l)}\big|\boldsymbol{c}\right)\big|U=l\right]\nonumber\\
    &\le \sum_{l\in\{F_{i_1}\cap F_{\tilde{i}_1}\}}\frac{1}{L}+2^{-n\delta}\\
    &=\frac{|F_{i_1}\cap F_{\tilde{i}_1}|}{L}+\circ(n).
    \label{eq.Pe2.result.cont}
\end{align*}
We define an auxiliary random variable $\Psi_l$ for all $l=1,\cdots,L$, where
\begin{align*}
    \Psi_l(F_{\tilde{i}_1})=\left\{
    \begin{array}{cc}
         1,& l\in F_{i_1}\cap F_{\tilde{i}_1}  \\
         0,& \text{otherwise}
    \end{array}
    \right.,
\end{align*}
with PMF
\begin{align*}
    Pr\left[\Psi_l=1\right]=\frac{1}{M_1}.
\end{align*}
Thus, for any fixed $i_1\ne \tilde{i}_1\in\mathcal{N}_1$, we can upper-bound the type II error by:
\begin{align*}
    P^1_{e,2}(i_1,\tilde{i_1})
    &\le \frac{|F_{i_1}\cap F_{\tilde{i}_1}|}{L}+\circ(n)\\
    &=\frac{1}{L}\sum_{l=1}^L\Psi_l(F_{\tilde{i}_1})+\circ(n).
\end{align*}
We introduce the following lemma.
\begin{lemma}\label{lemma:inequality}\cite{ahlswede1989identification} Let $\Psi_1,\cdots,\Psi_L$ be independent identically distributed RVs with value in $\{0,1\}$. Suppose that the expectation $\mathbb{E}\left[\Psi_1\right]\le\mu\le\lambda\le1$, then
\begin{align*}
    Pr\left[\sum_{l=1}^L\Psi_l>L\cdot \lambda \right]\le 2^{-L\cdot D(\lambda||\mu)},
\end{align*}
where $D(\lambda||\mu)$ denotes the information divergence between $(\lambda,1-\lambda)$ and $(\mu,1-\mu)$.
\end{lemma}
The following corollary is obtained by applying Lemma \ref{lemma:inequality}:
\begin{corollary}
    For $\lambda\in(0,1)$, and $\mathbb{E}\left[\Psi\right]=\frac{1}{M_1}\le\lambda$,
\begin{align*}
    Pr\left[\frac{1}{L}\sum_{l=1}^L\Psi_l(F_{\tilde{i}_1})>\lambda\right]<2^{-L\cdot(\lambda\log{M_1}-1)}.
\end{align*}
\end{corollary}

For every $\tilde{i}_1\neq i_1$, we should ensure that $P_{e,2}^1(i_1,\tilde{i}_1)\leq \lambda$. Therefore, we consider the probability of the following joint event:
\begin{align*}
    &Pr\left[ \bigcap_{\tilde{i}_1\ne i_1}\left\{\frac{1}{L}\sum_{l-1}^L\Psi_{l}(F_{\tilde{i}_1})\le\lambda\right\}\right]\\
    &=1-Pr\left[\bigcup_{\tilde{i}_1\ne i_1}\left\{\frac{1}{L}\sum_{l=1}^L\Psi_{l}(F_{\tilde{i}_1})>\lambda\right\}\right]\\
    &\ge1- \sum_{\tilde{i}_1\ne i_1} Pr\left[\frac{1}{L}\sum_{l=1}^L\Psi_{l}(F_{\tilde{i}_1})>\lambda\right]\nonumber\\
    &\ge1- (N_1-1)\cdot 2^{-L\cdot(\lambda\log{M_1}-1)}.
\end{align*}
We aim for the event $\bigcap_{\tilde{i}_1\ne i_1}\left\{\frac{1}{L}\sum_{l=1}^L\Psi_{l}(F_{\tilde{i}})\leq\lambda\right\}$ to hold with positive probability. Therefore, the maximum value of $N_1$ we can choose is
\begin{align*}
    N_1=2^{L\cdot(\lambda\log{M_1}-1)}.
\end{align*}
We choose $L$ to be any large integer satisfying 
\begin{align*}
    \lim_{n\to\infty}\frac{\log{\left(L\cdot(\lambda\log(M_1)-1)\right)}}{n}=+\infty,
\end{align*}
for example, we choose $L=2^{2^{n}}$, then
\begin{align*}
    \lim_{m\to\infty}R_1=\lim_{n\to\infty}\frac{\log\log{2^{\left(2^{2^{n}}\cdot(\lambda\log(M_1)-1)\right)}}}{n+1}=+\infty.
\end{align*}
Similarly, for all $k\in\mathcal{K}$, we can achieve
\begin{align*}
    \lim_{m\to\infty}R_k=+\infty.
\end{align*}
This completes the proof of Theorem \ref{thm:GMAC}.
\subsection{Sketch proof of Theorem \ref{thm:SD-GMAC}}
Similarly, consider an IDF code $\left(m,\boldsymbol{N}=2^{2^{m\boldsymbol{R}}},\lambda\right)$ with code length $m=n+1$. Each sender $k\in\mathcal{K}$ firstly sends $x_{k,1}=0$. According to our assumptions, $\boldsymbol{S}\sim \mathcal{N}\left(\boldsymbol{\mu},\boldsymbol{\Sigma}\right)$ and $Z\sim \mathcal{N}(0,\sigma^2)$, and they are independent. Therefore, we have $Y=\sum_{k=1}^KS_{k}+Z\sim \mathcal{N}\left(\mu_Y,\sigma_Y^2\right)$. 
We utilize the same code construction scheme as previously discussed.

Next, we examine the type I and type II errors. We define the following average channel $W^a_{\sigma^2,S}$, averaging over all channel states:
\begin{align*}
    W^a_{\sigma^2,S}(y|\boldsymbol{x})={\int_{\boldsymbol{s}}} p_{\boldsymbol{S}}(\boldsymbol{s})W_{\sigma^2,S}(y|\boldsymbol{x},\boldsymbol{s})d\boldsymbol{s}.
\end{align*}
For all $\boldsymbol{i}\in\boldsymbol{\mathcal{N}}$, the type I error can be written as
\begin{align*}
    P_{e,1}\left(\boldsymbol{i}\right)=W^a_{\sigma^2,S}\left(\mathcal{D}^c_{\boldsymbol{i}}|\boldsymbol{f}_{\boldsymbol{i}}\right),
\end{align*}
and for all $\boldsymbol{i},\boldsymbol{\tilde{i}}\in\boldsymbol{\mathcal{N}}$, $\boldsymbol{i}\ne \boldsymbol{\tilde{i}}$, the type II error can be written as
\begin{align*}
    P_{e,2}\left(\boldsymbol{i},\tilde{\boldsymbol{i}}\right)=W^a_{\sigma^2,S}(\mathcal{D}_{\boldsymbol{\tilde{i}}}|\boldsymbol{f}_{\boldsymbol{i}}).
\end{align*}
The subsequent proof is identical to the previous proof of Theorem \ref{thm:GMAC}.

\section{Discussion and Conclusions}
\label{sec:conclusion}
In this section, we summarize the findings about the scaling function used
in the previous studies on transmission and ID capacity and explore the capacity region of K-GMAC when altering the scaling of rates. We begin by introducing the definition of the scaling function.
\begin{definition}
\label{def.scaling}
    \cite{labidi2022common} Let $\varphi: \mathbb{R}^+\mapsto \mathbb{R}^+$ with $\lim_{n\to\infty} \varphi(nR)=+\infty$ be an arbitrary continuous strictly monotonically increasing function, which quantifies the relationship between the code size $N$ and the block length $n$, i.e., i.e., $N=\varphi(nR)$.

\end{definition}
We choose the scaling function under which the rate is positive but finite. Commonly used scaling functions for transmission and ID rates include $\varphi_1(nR)=2^{nR}$, $\varphi_2(nR)=n^{nR}$, and $\varphi_3(nR)=2^{2^{nR}}$. The transmission rate is typically defined w.r.t. $\varphi_1$\cite{shannon1948mathematical}. For ID via DMCs without feedback, the deterministic ID rate is defined w.r.t. $\varphi_1$ \cite{salariseddigh2021deterministicpower,796419}, while the randomized ID rate is defined w.r.t. $\varphi_3$ \cite{ahlswede1995new}. With perfect feedback, both deterministic and randomized IDF rates are defined w.r.t. $\varphi_3$ \cite{ahlswede1989identificationfeedback}. In the case of continuous channels, the deterministic ID rate is defined w.r.t. $\varphi_2$ \cite{salariseddigh2023deterministic,salariseddigh2021deterministicfading,inproceedings}. A "slower" scaling leads to an infinite rate, while a "faster" scaling results in a zero rate. For instance, deterministic ID rate of Gaussian channels yields infinite rate with $\varphi_1$ and zero rate with $\varphi_3$, i.e., $R^{\varphi_1}_{dID}(W_{\sigma^2})=\frac{\log{N}}{n}= +\infty$ and $R^{\varphi_3}_{dID}(W_{\sigma^2})=\frac{\log{\log{N}}}{n}= 0$. However, it has been shown that the IDF capacity of single-user continuous channels with additive noise is infinite regardless of the scaling function used \cite{9716126}. For our IDF problem via K-GMAC and SD-K-GMAC, we have the following corollary.
\begin{corollary}\label{corollary:scaling} If $P_{total}>0$, regardless of the scaling function $\varphi$ used, each sender of K-GAMC or K-SD-GMAC can achieve an infinite rate under average power constraint parameter $P$.
\end{corollary}
This can be easily demonstrated by selecting $L$ to satisfy
\begin{align*}
    \lim_{n\to\infty}\frac{\varphi^{-1}\left(2^{L\cdot (\lambda\log{M_k}-1)}\right)}{n}=+\infty,\quad \forall k\in\mathcal{K}.
\end{align*}

In conclusion, our investigation into the IDF capacity region under power constraints of K-GMAC and K-SD-GMAC with additive Gaussian channel states reveals that irrespective of the scaling of rate, each sender can achieve an infinite rate by effectively partitioning the feedback into an appropriate number of parts. A potential research direction is to extend the capacity region of GMAC to MAC with any non-discrete additive noise. However, it is noteworthy that in real communication systems, the decoder or the feedback might be rate-limited \cite{vahid2011interference,ardestanizadeh2009wiretap}. If at least one of them is rate-limited, then the IDF capacity of each sender is no longer infinite.
\section*{Acknowledgments}
H. Boche, C. Deppe, W. Labidi and Y. Zhao acknowledge the financial support by the Federal Ministry of Education and Research
of Germany (BMBF) in the programme of “Souverän. Digital. Vernetzt.”. Joint project 6G-life, project identification number: 16KISK002.
Eduard A. Jorswieck is supported by
the Federal Ministry of Education and Research (BMBF, Germany) through
the Program of “Souveran. Digital. Vernetzt.” Joint Project 6G-Research and
Innovation Cluster (6G-RIC) under Grant 16KISK031. 
H. Boche and W. Labidi were further supported in part by the BMBF within the national initiative on Post Shannon Communication (NewCom) under Grant 16KIS1003K. C.\ Deppe was further supported in part by the BMBF within NewCom under Grant 16KIS1005. C. Deppe, W. Labidi and Y. Zhao were also supported by the DFG within the project DE1915/2-1. 
\bibliographystyle{IEEEtran}
\bibliography{definitions,references}

\IEEEtriggeratref{4}



\end{document}

%% file: Figures/IDFsingle.tex
\scalebox{1}{
\tikzstyle{block} = [draw, fill=white, rectangle, minimum height=3em, minimum width=4.5em,scale=0.7]
    \tikzstyle{input} = [coordinate]
    \tikzstyle{pinstyle} = [pin edge={to-,thin,black}]
    \tikzstyle{point} = [draw, fill, circle, inner sep=0.5pt, outer sep=0pt]
    \tikzstyle{smallblock}=[draw,fill=white, rectangle, minimum height =0.8em, minimum width=0.8em]
    \tikzstyle{channel}=[draw, fill=white, circle, inner sep=1pt, outer sep=0pt]
    \centering
    \begin{tikzpicture}[auto, node distance=3cm]
        \node[input](source){Source};
        \node[block,right of=source,node distance=2.2cm](encoder){Encoder};
        \node[channel,right of=encoder,node distance=3cm](channel){\Large \bf $+$};
        \node[input,above of=channel,node distance=0.8cm](noise){};
        \node[above of =noise,node distance=0.15cm]{\small $Z_t$};
        \node[block,right of=channel,node distance=3cm](decoder){Decoder};
        \node[input,below of=decoder,node distance=0.39cm,pin={[pinstyle]below:\scriptsize{$j\in\mathcal{N}$}}](i'){};
        \node[input,right of=decoder,node distance=1.8cm](sink){Sink};
        \node[point,right of= channel,node distance=1.3cm](p1){};
        \node[smallblock,below of=p1,node distance=1cm](p2){\small D};
        \node[input,below of=encoder,node distance=1cm](p3){};
        \draw [->] (source) -- node[name=m] {\scriptsize$i\in\mathcal{N}$} (encoder);
        \draw[->](encoder) -- node[name=x] {\scriptsize$X_t=f_i(Y^{t-1})$}(channel);
        \draw[->](channel) -- node[name=y]{\scriptsize$Y_t\quad$}(decoder);
        \draw[->](decoder) -- node[]{\scriptsize$i=j?$}(sink);
        \draw[-](p1)--(p2);
        \draw[-](p2)--node[above]{\small $Y_{t-1}$}(p3);
        \draw[->](p3)--(encoder);
        \draw[->](noise)--(channel);
    \end{tikzpicture}
    }

%% file: Figures/GAMC.tex
\scalebox{0.68}{
\tikzstyle{block} = [draw, fill=white, rectangle, minimum height=3em, minimum width=4em]
    \tikzstyle{lblock}=[draw, fill=white, rectangle, minimum height=8.5em, minimum width=4em]
    \tikzstyle{channel}=[draw, fill=white, circle, inner sep=1pt, outer sep=0pt]
    \tikzstyle{point} = [draw, fill, circle, inner sep=1pt, outer sep=0pt]
    \tikzstyle{input} = [coordinate]
    \tikzstyle{output} = [coordinate]
    \tikzstyle{pinstyle} = [pin edge={to-,thin,black}]
    \tikzstyle{smallblock}=[draw,fill=white, rectangle, minimum height =2em, minimum width=2em]
    \tikzstyle{box}=[draw, rectangle,minimum height=6.5em, minimum width=6.7cm]
    \begin{tikzpicture}[auto, node distance=3cm]
        \node[input](source1){Source1};
        \node[block,right of=source1,node distance=2.5cm](encoder1){Encoder 1};
        \node[input,below of=source1,node distance=2cm](source2){Source2};
        \node[block,below of=encoder1,node distance=2cm](encoder2){Encoder K};
        \node[input,below of=encoder1,node distance=1cm](input){};
        \node[channel,right of=input,node distance=5cm,align=center](channel){\LARGE{\textbf{$+$}}};
        \node[input,right of=encoder1,node distance=4cm](channel_input_1){};
        \node[input,right of=encoder2,node distance=4cm](channel_input_2){};
        \node[lblock,right of=channel,node distance=2.8cm](decoder){Decoder};
        \node[input,align=center,below of=decoder,node distance=2cm](i'j'){};
        \node[output,right of=decoder, node distance=2.5cm](sink){Sink};
        \node[input,below of=channel,node distance=0.45cm,pin={[pinstyle]below:$Z_t$}](noise){};
    
        \node[point,right of=channel,node distance=1.6cm](f1){};
        \node[smallblock,above of=f1,node distance=2cm](f2){D};
        \node[input,above of=encoder1,node distance=1cm](f3){};
        \node[input,above of=encoder1,node distance=0.6cm](f4){};
        \node[smallblock,below of=f1,node distance=2.3cm](f5){D};
        \node[input,below of=encoder2,node distance=1.3cm](f6){};
        \node[input,below of=encoder2,node distance=0.6cm](f7){}; 
        \draw [->] (source1) -- node[name=i] {$i_1\in\mathcal{N}_1$} (encoder1);
        \draw[->](source2) --  node[name=j] {$i_K\in\mathcal{N}_K$} (encoder2);
        \draw[-](encoder1) -- node[name=x1,align=right] {$X_{1,t}=f_{1,i_1}^t\left(Y^{t-1}\right)$}(channel_input_1);
        \draw[->](channel_input_1)--(channel);
        \draw[-](encoder2) -- node[name=x2,align=right] {$X_{K,t}=f_{K,i_K}^t\left(Y^{t-1}\right)$}(channel_input_2);
        \draw[->](channel_input_2)--(channel);
        \draw[->](channel) -- node[name=y]{$Y_t\quad$}(decoder);
        \draw[->](decoder) -- node[name=result,align=center]{$i_1=j_1?$\\$\cdots$\\$i_K=j_K?$}(sink);
        \draw[-](f1)--(f2);
        \draw[-](f2)--node[above]{$Y_{t-1}$}(f3);
        \draw[->](f3)--(f4);
        \draw[-](f1)--(f5);
        \draw[-](f5)--node[above]{$Y_{t-1}$}(f6);
        \draw[->](f6)--(f7);
        \draw[->](i'j')--(decoder);
        \node[] at (0.8,-1) {\rotatebox{-90}{\textbf{$\cdots$}}};
        \node[] at (2.5,-1) {\rotatebox{-90}{\textbf{$\cdots$}}};
        \node[] at (4.2,-1) {\rotatebox{-90}{\textbf{$\cdots$}}};
        \node[below of = decoder, node distance = 2.8cm,align=center]{$j_1\in\mathcal{N}_1$\\$\cdots$\\$j_K\in\mathcal{N}_K$};
       \node[box, dashed,gray] at (3.2,0.5) (sender) {};
       \node[box, dashed,gray] at (3.2,-2.5) (sender) {};
       \node[] at (0.7,1.3) () {Sender 1};
       \node[] at (0.7,-3.2) () {Sender K};
    \end{tikzpicture}
    }

%% file: Figures/SDGMAC.tex
\scalebox{0.68}{
\tikzstyle{block} = [draw, fill=white, rectangle, minimum height=3em, minimum width=4em]
    \tikzstyle{lblock}=[draw, fill=white, rectangle, minimum height=8.5em, minimum width=4em]
    \tikzstyle{channel}=[draw, fill=white, circle, inner sep=1pt, outer sep=0pt]
    \tikzstyle{point} = [draw, fill, circle, inner sep=1pt, outer sep=0pt]
    \tikzstyle{input} = [coordinate]
    \tikzstyle{output} = [coordinate]
    \tikzstyle{pinstyle} = [pin edge={to-,thin,black}]
    \tikzstyle{smallblock}=[draw,fill=white, rectangle, minimum height =2em, minimum width=2em]
    \tikzstyle{box}=[draw, rectangle,minimum height=6.8em, minimum width=5cm]
    \begin{tikzpicture}[auto, node distance=3cm]
        \node[input](source1){Source1};
        \node[block,right of=source1,node distance=2.8cm](encoder1){Encoder 1};
        \node[input,below of=source1,node distance=2cm](source2){Source2};
        \node[block,below of=encoder1,node distance=2cm](encoder2){Encoder K};
        \node[input,below of=encoder1,node distance=1cm](input){};
        \node[channel,right of=input,node distance=4.3cm,align=center](channel){\Large\textbf{$+$}};
        \node[channel,right of=encoder1, node distance =2.5cm](state1){\Large\textbf{$+$}};
        \node[input,right of=state1,node distance=2cm](channel_input_1){};
        \node[channel,right of=encoder2,node distance=2.5cm](state2){\Large\textbf{$+$}};
        \node[input,right of=state2,node distance=2cm](channel_input_2){};
        \node[lblock,right of=channel,node distance=2.8cm](decoder){Decoder};
        \node[input,below of=decoder,node distance=2cm](i'j'){};
        \node[input,right of=decoder,node distance=2.5cm](sink){Sink};
        \node[input,below of=channel,node distance=0.4cm,pin={[pinstyle]below:$Z_t$}](noise){};
        \node[input,above of=state1,node distance=0.35cm,pin={[pinstyle]above:$S_{1,t}$}](s1){};
        \node[input,below of=state2,node distance=0.35cm,pin={[pinstyle]below:$S_{K,t}$}](s1){};
        \node[point,right of=channel,node distance=1.6cm](f1){};
        \node[smallblock,above of=f1,node distance=2.6cm](f2){D};
        \node[input,above of=encoder1,node distance=1.6cm](f3){};
        \node[input,above of=encoder1,node distance=0.6cm](f4){};
        \node[smallblock,below of=f1,node distance=2.6cm](f5){D};
        \node[input,below of=encoder2,node distance=1.6cm](f6){};
        \node[input,below of=encoder2,node distance=0.6cm](f7){}; 
        \draw [->] (source1) -- node[name=i] {$i_1\in\mathcal{N}_1$} (encoder1);
        \draw[->](source2) --  node[name=j] {$i_K\in\mathcal{N}_K$} (encoder2);
        \draw[->](encoder1) -- node[name=x1,align=right] {$X_{1,t}\quad$}(state1);
        \draw[->](encoder2) -- node[name=x2,align=right] {$X_{K,t}\quad$}(state2);
        \draw[->](channel) -- node[name=y]{$Y_t\quad$}(decoder);
        \draw[->](decoder) -- node[name=result,align=center]{$i_1=j_1?$\\$\cdots$\\$i_K=j_K?$}(sink);
        \draw[-](f1)--(f2);
        \draw[-](f2)--(f3);
        \draw[->](f3)--node[left]{$Y_{t-1}$}(f4);
        \draw[-](f1)--(f5);
        \draw[-](f5)--(f6);
        \draw[->](f6)--node[left]{$Y_{t-1}$}(f7);
        \draw[->](state1)--(channel);
        \draw[->](state2)--(channel);
        \draw[->](i'j')--(decoder);
        \node[below of = decoder, node distance = 2.8cm,align=center]{$j_1\in\mathcal{N}_1$\\$\cdots$\\$j_K\in\mathcal{N}_K$};
        \node[] at (0.8,-1) {\rotatebox{-90}{\textbf{$\cdots$}}};
        \node[] at (2.8,-1) {\rotatebox{-90}{\textbf{$\cdots$}}};
        \node[] at (4.4,-1) {\rotatebox{-90}{\textbf{$\cdots$}}};
        \node[box, dashed,gray] at (2.2,0.6) (sender) {};
        \node[box, dashed,gray] at (2.2,-2.6) (sender) {};
        \node[] at (0.6,1.5) () {Sender 1};
        \node[] at (0.6,-3.4) () {Sender K};
    \end{tikzpicture}
    }

%% file: arxiv.bbl
\begin{thebibliography}{10}
\providecommand{\url}[1]{#1}
\csname url@samestyle\endcsname
\providecommand{\newblock}{\relax}
\providecommand{\bibinfo}[2]{#2}
\providecommand{\BIBentrySTDinterwordspacing}{\spaceskip=0pt\relax}
\providecommand{\BIBentryALTinterwordstretchfactor}{4}
\providecommand{\BIBentryALTinterwordspacing}{\spaceskip=\fontdimen2\font plus
\BIBentryALTinterwordstretchfactor\fontdimen3\font minus
  \fontdimen4\font\relax}
\providecommand{\BIBforeignlanguage}[2]{{%
\expandafter\ifx\csname l@#1\endcsname\relax
\typeout{** WARNING: IEEEtran.bst: No hyphenation pattern has been}%
\typeout{** loaded for the language `#1'. Using the pattern for}%
\typeout{** the default language instead.}%
\else
\language=\csname l@#1\endcsname
\fi
#2}}
\providecommand{\BIBdecl}{\relax}
\BIBdecl

\bibitem{shannon1948mathematical}
C.~E. Shannon, ``A mathematical theory of communication,'' \emph{The Bell
  system technical journal}, vol.~27, no.~3, pp. 379--423, 1948.

\bibitem{fettweis20226g}
G.~P. Fettweis and H.~Boche, ``On {6G} and trustworthiness,''
  \emph{Communications of the ACM}, vol.~65, no.~4, pp. 48--49, 2022.

\bibitem{cabrera20216g}
J.~A. Cabrera, H.~Boche, C.~Deppe, R.~F. Schaefer, C.~Scheunert, and F.~H.
  Fitzek, ``{6G} and the post-shannon theory,'' \emph{Shaping Future {6G}
  Networks: Needs, Impacts, and Technologies}, pp. 271--294, 2021.

\bibitem{boche2018secure}
H.~Boche and C.~Deppe, ``Secure identification for wiretap channels;
  robustness, super-additivity and continuity,'' \emph{IEEE Transactions on
  Information Forensics and Security}, vol.~13, no.~7, pp. 1641--1655, 2018.

\bibitem{moulin2001role}
P.~Moulin, ``The role of information theory in watermarking and its application
  to image watermarking,'' \emph{Signal Processing}, vol.~81, no.~6, pp.
  1121--1139, 2001.

\bibitem{ahlswede2006watermarking}
R.~Ahlswede and N.~Cai, ``Watermarking identification codes with related topics
  on common randomness,'' \emph{General Theory of Information Transfer and
  Combinatorics}, pp. 107--153, 2006.

\bibitem{steinberg2001identification}
Y.~Steinberg and N.~Merhav, ``Identification in the presence of side
  information with application to watermarking,'' \emph{IEEE Transactions on
  Information Theory}, vol.~47, no.~4, pp. 1410--1422, 2001.

\bibitem{lu2017industry}
Y.~Lu, ``Industry 4.0: A survey on technologies, applications and open research
  issues,'' \emph{Journal of industrial information integration}, vol.~6, pp.
  1--10, 2017.

\bibitem{Schwenteck2023}
P.~Schwenteck, G.~T. Nguyen, H.~Boche, W.~Kellerer, and F.~H.~P. Fitzek, ``6g
  perspective of mobile network operators, manufacturers, and verticals,''
  \emph{{IEEE} Network Letters}, vol.~5, no.~3, pp. 169--172, 2023.

\bibitem{ahlswede1989identification}
R.~Ahlswede and G.~Dueck, ``Identification via channels,'' \emph{IEEE
  Transactions on Information Theory}, vol.~35, no.~1, pp. 15--29, 1989.

\bibitem{ja1985identification}
J.~JaJa, ``Identification is easier than decoding,'' in \emph{26th Annual
  Symposium on Foundations of Computer Science (sfcs 1985)}.\hskip 1em plus
  0.5em minus 0.4em\relax IEEE, 1985, pp. 43--50.

\bibitem{ahlswede1995new}
R.~Ahlswede and Z.~Zhang, ``New directions in the theory of identification via
  channels,'' \emph{IEEE transactions on information theory}, vol.~41, no.~4,
  pp. 1040--1050, 1995.

\bibitem{labidi2020secure}
W.~Labidi, C.~Deppe, and H.~Boche, ``Secure identification for {Gaussian}
  channels,'' in \emph{ICASSP 2020-2020 IEEE International Conference on
  Acoustics, Speech and Signal Processing (ICASSP)}.\hskip 1em plus 0.5em minus
  0.4em\relax IEEE, 2020, pp. 2872--2876.

\bibitem{boche2019secure}
H.~Boche, C.~Deppe, and A.~Winter, ``Secure and robust identification via
  classical-quantum channels,'' \emph{IEEE Transactions on Information Theory},
  vol.~65, no.~10, pp. 6734--6749, 2019.

\bibitem{pereg2022identification}
U.~Pereg, J.~Rosenberger, and C.~Deppe, ``Identification over quantum broadcast
  channels,'' in \emph{2022 IEEE International Symposium on Information Theory
  (ISIT)}.\hskip 1em plus 0.5em minus 0.4em\relax IEEE, 2022, pp. 258--263.

\bibitem{ezzine2024common}
R.~Ezzine, M.~Wiese, C.~Deppe, and H.~Boche, ``Common randomness generation
  from finite compound sources,'' in \emph{2024 {IEEE} International Symposium
  on Information Theory ({ISIT})}.\hskip 1em plus 0.5em minus 0.4em\relax
  {IEEE}, 2024.

\bibitem{ezzine2021common}
------, ``Common randomness generation over slow fading channels,'' in
  \emph{2021 IEEE International Symposium on Information Theory (ISIT)}.\hskip
  1em plus 0.5em minus 0.4em\relax IEEE, 2021, pp. 1925--1930.

\bibitem{labidi2022common}
W.~Labidi, R.~Ezzine, C.~Deppe, and H.~Boche, ``Common randomness generation
  from gaussian sources,'' in \emph{2022 IEEE International Symposium on
  Information Theory (ISIT)}, 2022, pp. 1548--1553.

\bibitem{ahlswede1989identificationfeedback}
R.~Ahlswede and G.~Dueck, ``Identification in the presence of feedback-a
  discovery of new capacity formulas,'' \emph{IEEE Transactions on Information
  Theory}, vol.~35, no.~1, pp. 30--36, 1989.

\bibitem{wolfowitz2012coding}
J.~Wolfowitz, \emph{Coding theorems of information theory}.\hskip 1em plus
  0.5em minus 0.4em\relax Springer Science \& Business Media, 2012, vol.~31.

\bibitem{cover1989gaussian}
T.~M. Cover and S.~Pombra, ``{Gaussian} feedback capacity,'' \emph{IEEE
  Transactions on Information Theory}, vol.~35, no.~1, pp. 37--43, 1989.

\bibitem{koga2002information}
H.~Koga \emph{et~al.}, \emph{Information-spectrum methods in information
  theory}.\hskip 1em plus 0.5em minus 0.4em\relax Springer Science \& Business
  Media, 2002, vol.~50.

\bibitem{burnashev1999method}
M.~V. Burnashev, ``On method of" types", approximation of output measures and
  id-capacity for channels with continuous alphabets,'' in \emph{Proceedings of
  the 1999 IEEE Information Theory and Communications Workshop (Cat. No.
  99EX253)}.\hskip 1em plus 0.5em minus 0.4em\relax IEEE, 1999, pp. 80--81.

\bibitem{salariseddigh2021deterministicfading}
M.~J. Salariseddigh, U.~Pereg, H.~Boche, and C.~Deppe, ``Deterministic
  identification over fading channels,'' in \emph{2020 IEEE Information Theory
  Workshop (ITW)}.\hskip 1em plus 0.5em minus 0.4em\relax IEEE, 2021, pp. 1--5.

\bibitem{ezzine2020common}
R.~Ezzine, W.~Labidi, H.~Boche, and C.~Deppe, ``Common randomness generation
  and identification over {Gaussian} channels,'' in \emph{GLOBECOM 2020-2020
  IEEE Global Communications Conference}.\hskip 1em plus 0.5em minus
  0.4em\relax IEEE, 2020, pp. 1--6.

\bibitem{labidi2021identification}
W.~Labidi, H.~Boche, C.~Deppe, and M.~Wiese, ``Identification over the
  {Gaussian} channel in the presence of feedback,'' in \emph{2021 IEEE
  International Symposium on Information Theory (ISIT)}.\hskip 1em plus 0.5em
  minus 0.4em\relax IEEE, 2021, pp. 278--283.

\bibitem{9716126}
M.~Wiese, W.~Labidi, C.~Deppe, and H.~Boche, ``Identification over additive
  noise channels in the presence of feedback,'' \emph{IEEE Transactions on
  Information Theory}, vol.~69, no.~11, pp. 6811--6821, 2023.

\bibitem{liao1972multiple}
H.~H.-J. Liao, ``Multiple access channels,'' Ph.D. dissertation, University of
  Hawaii Honolulu, HI, USA, 1972.

\bibitem{el2011network}
A.~El~Gamal and Y.-H. Kim, \emph{Network information theory}.\hskip 1em plus
  0.5em minus 0.4em\relax Cambridge university press, 2011.

\bibitem{gaarder1975capacity}
N.~Gaarder and J.~Wolf, ``The capacity region of a multiple-access discrete
  memoryless channel can increase with feedback (corresp.),'' \emph{IEEE
  Transactions on Information Theory}, vol.~21, no.~1, pp. 100--102, 1975.

\bibitem{cover1999elements}
T.~M. Cover, \emph{Elements of information theory}.\hskip 1em plus 0.5em minus
  0.4em\relax John Wiley \& Sons, 1999.

\bibitem{thomas1987feedback}
J.~Thomas, ``Feedback can at most double {Gaussian} multiple access channel
  capacity (corresp.),'' \emph{IEEE transactions on Information theory},
  vol.~33, no.~5, pp. 711--716, 1987.

\bibitem{rosenberger2023deterministic}
J.~Rosenberger, A.~Ibrahim, C.~Deppe, and R.~Ferrara, ``Deterministic
  identification over multiple-access channels,'' in \emph{2023 IEEE
  International Symposium on Information Theory (ISIT)}, 2023.

\bibitem{ahlswede2008general}
R.~Ahlswede, ``General theory of information transfer: Updated,''
  \emph{Discrete Applied Mathematics}, vol. 156, no.~9, pp. 1348--1388, 2008.

\bibitem{diadamo2019simultaneous}
\BIBentryALTinterwordspacing
S.~Diadamo and H.~Boche, ``The simultaneous identification capacity of the
  classical--quantum multiple access channel with stochastic encoders for
  transmission,'' 2019. [Online]. Available:
  \url{https://arxiv.org/abs/1903.03395}
\BIBentrySTDinterwordspacing

\bibitem{ahlswede1971multi}
R.~Ahlswede, ``Multi-way communication channels,'' in \emph{Proc. 2nd. Int.
  Symp. Information Theory (Tsahkadsor, Armenian SSR), 1971}.\hskip 1em plus
  0.5em minus 0.4em\relax Publishing House of the Hungarian Academy of
  Sciences, 1971, pp. 23--52.

\bibitem{Kleinewachter1999}
C.~Kleinewachter, ``Identification via noiseless channels with noisy
  feedback,'' in \emph{Proceedings of the 1999 {IEEE} Information Theory and
  Communications Workshop (Cat. No. 99EX253)}, 1999, p.~84.

\bibitem{salariseddigh2021deterministicpower}
M.~J. Salariseddigh, U.~Pereg, H.~Boche, and C.~Deppe, ``Deterministic
  identification over channels with power constraints,'' \emph{IEEE
  Transactions on Information Theory}, vol.~68, no.~1, pp. 1--24, 2021.

\bibitem{796419}
R.~Ahlswede and N.~Cai, ``Identification without randomization,'' \emph{IEEE
  Transactions on Information Theory}, vol.~45, no.~7, pp. 2636--2642, 1999.

\bibitem{salariseddigh2023deterministic}
M.~J. Salariseddigh, V.~Jamali, U.~Pereg, H.~Boche, C.~Deppe, and R.~Schober,
  ``Deterministic identification for mc isi-poisson channel,'' in \emph{ICC
  2023-IEEE International Conference on Communications}.\hskip 1em plus 0.5em
  minus 0.4em\relax IEEE, 2023, pp. 6108--6113.

\bibitem{inproceedings}
M.~J. Salariseddigh, U.~Pereg, H.~Boche, C.~Deppe, and R.~Schober,
  ``Deterministic identification over poisson channels,'' 12 2021, pp. 1--6.

\bibitem{vahid2011interference}
A.~Vahid, C.~Suh, and A.~S. Avestimehr, ``Interference channels with
  rate-limited feedback,'' \emph{IEEE Transactions on Information Theory},
  vol.~58, no.~5, pp. 2788--2812, 2011.

\bibitem{ardestanizadeh2009wiretap}
E.~Ardestanizadeh, M.~Franceschetti, T.~Javidi, and Y.-H. Kim, ``Wiretap
  channel with secure rate-limited feedback,'' \emph{IEEE Transactions on
  Information Theory}, vol.~55, no.~12, pp. 5353--5361, 2009.

\end{thebibliography}
